\documentclass{article}

\usepackage[nonatbib,preprint]{neurips_2022}

\usepackage[utf8]{inputenc} 
\usepackage[T1]{fontenc}    
\usepackage{hyperref}       
\usepackage{url}            
\usepackage{booktabs}       
\usepackage{amsfonts}       
\usepackage{nicefrac}       
\usepackage{microtype}      
\usepackage{xcolor}         
\usepackage[linesnumbered,boxed,ruled,commentsnumbered]{algorithm2e}
\usepackage{amsmath}
\usepackage{amssymb}
\usepackage{graphicx}
\usepackage{threeparttable}
\usepackage{multirow}
\usepackage{subfigure}
\usepackage{floatrow,caption}

\title{Fed-FSNet: Mitigating Non-I.I.D. Federated Learning \\
via Fuzzy Synthesizing Network}

\author{%
  Jingcai Guo, Song Guo, Jie Zhang, and Ziming Liu \\
  Department of Computing, The Hong Kong Polytechnic University \\
  \texttt{\{jingcai.guo,jieaa.zhang,ziming.liu\}@connect.polyu.hk,song.guo@polyu.edu.hk} \\
}

\begin{document}

\maketitle

\begin{abstract}
Federated learning (FL) has emerged as a promising privacy-preserving distributed machine learning framework recently. It aims at collaboratively learning a shared global model by performing distributed training locally on edge devices and aggregating local models into a global one without centralized raw data sharing in the cloud server. However, due to the large local \textit{data heterogeneities} (Non-I.I.D. data) across edge devices, the FL may easily obtain a global model that can produce more \textit{shifted gradients} on local datasets, thereby degrading the model performance or even suffering from the non-convergence during training. In this paper, we propose a novel FL training framework, dubbed \textit{Fed-FSNet}, using a properly designed \underline{F}uzzy \underline{S}ynthesizing \underline{Net}work (FSNet) to mitigate the Non-I.I.D. FL \textit{at-the-source}. Concretely, we maintain an \textit{edge-agnostic} hidden model in the cloud server to estimate a less-accurate while \textit{direction-aware} inversion of the global model. The hidden model can then fuzzily synthesize several mimic I.I.D. data samples (sample features) conditioned on only the global model, which can be shared by edge devices to facilitate the FL training towards faster and better convergence. Moreover, since the synthesizing process involves neither access to the parameters/updates of local models nor analyzing individual local model outputs, our framework can still \textit{ensure the privacy of FL}. Experimental results on several FL benchmarks demonstrate that our method can significantly mitigate the Non-I.I.D. issue and obtain better performance against other representative methods.
\end{abstract}

\section{Introduction}
Machine deep learning has made tremendous success in the past few years across multiple real-world applications~\cite{deng2009imagenet,guo2020novel,lv2014traffic,nguyen2015topic,guo2020dual,lian2018xdeepfm,simonyan2014very,guo2019adaptive,yang2015facial,shone2018deep,guo2023graph,liu2022towards,guo2021conservative,lu2023decomposed,wang2023data,liu2023zsl,guo2023application,huo2023offline,zhou2021device,guo2021learning,wang2022exploring,ma2019position,guo2016improved,huo2022procc,zhou2022cadm,guo2019ams,wang2022efficient,guo2019ee,zhou2021octo,guo2022fed}. 
With the advances in computing capability of edge devices (e.g., mobile phones), various machine learning (ML) applications~\cite{guo2016improved,ren2019fastspeech,anwar2019real,guo2020dual,kirillov2020pointrend,guo2020novel,balashankar2021learning,zhang2021cross,guo2021conservative,liu2022towards} are usually deployed locally on user-sides to directly perform training or inference, rather than relying on a centralized cloud server-based training framework for better privacy preservation \cite{xie2019slsgd,ochiai2019real,ravi2019efficient,bistritz2020distributed,zhou2021device,xian2021communication,zhou2021octo}. More specifically, users can either train an ML model individually upon their local data or download a pre-trained learned from external data from the cloud server. However, under such a paradigm, the former scenario may usually be quite stuck with local devices because the models cannot generalize well to massive new data, while the latter may result in a too mediocre model for each user by using only external data resources. 
For example, as one of the most ubiquitous edge devices in our daily life, the mobile phone can constantly generate privacy-sensitive data across multiple users, such as images, texts, voices, and so on. These data can represent sufficient diversity of various user patterns and thus are of great value to be utilized to train a better ML model for all participated users. However, due to privacy concerns, it is not applicable to upload and share these data in the cloud server for centralized model training, where the limitation is considered as the \textit{isolated-data-island} that hinders the further improvement of the model \cite{yang2019federated}. 
To facilitate the usage of such privacy-sensitive data and training a more generalized model, federated learning (FL) has emerged as a promising paradigm of privacy-preserving distributed ML framework \cite{mcmahan2017communication,smith2017federated,brisimi2018federated,bhagoji2019analyzing,ng2020multi,wang2020optimizing,lin2020ensemble, tang2022personalized}. 
FL enables each device to train a model locally based on its own data, and communicates with other local models in the cloud server for aggregating a more generalized global model, without collecting and sharing any privacy-sensitive information from users. 

However, unlike the standard cloud server-based distributed ML training framework, the FL is fundamentally posed with some challenges that limit its further development. 
Among them, due to the large heterogeneities across multiple devices, their data samples are usually not independent and identically distributed (Non-I.I.D.) \cite{hsieh2020non,castano2001global}. In such a scenario, the conventional FL is not able to efficiently train a shared global model and may thus degrade the performance or even encounter non-convergence during training. 
To mitigate the Non-I.I.D. FL, existing methods mostly follow two mainstreams. The first usually considers properly sampling several devices (or local training data) to participate in each round of FL to counterbalance the bias introduced by non-I.I.D. data \cite{mcmahan2017communication,goetz2019active,wang2020optimizing}. Such treatment is based on the assumption that there may exist (partially) I.I.D. cases across devices. 
The second considers exploring a balanced aggregation strategy for each local model (or model updates) to offset the Non-I.I.D. issue \cite{li2018federated,ji2019learning,lin2020ensemble,zhu2021data}, of which each device is assumed to contribute differently to the overall model. 
Despite the progress made, these methods can only make use of limited statistical observations associated with local devices, e.g., local model parameters/updates or sketchy data distributions, which are not sufficient to address the data Non-I.I.D. issue \textit{at-the-source}. 
Worse still, unlike the secure aggregation-based methods \cite{bonawitz2017practical,so2021turbo} where the cloud server can only observe the \textit{final sum} of model parameters/updates, the above-mentioned methods assign the priority of observing the parameters/updates of \textit{each local model} to the cloud server, thereby making the FL training not fully privacy-preserved if the cloud server is adversarial-oriented and may potentially reconstruct raw data of local devices. 

In this paper, we consider a novel strategy to synthesize mimic I.I.D. data samples (sample features) in a global view and offload them to participating local devices to directly mitigate the Non-I.I.D. FL \textit{at-the-source}. 
Specifically, we propose a well-designed \underline{F}uzzy \underline{S}ynthesizing \underline{Net}work (\textit{FSNet}) that maintains an \textit{edge-agnostic} hidden model in the cloud server, by using an encoder-decoder architecture associated with the global model to estimate an inversion of it. This inversion is typically achieved by sampling Gaussian mixture with variational mixture coefficients, to perform encoding and reconstruction process by the global and the hidden model, respectively. Notably, since the sampling is not equivalent to the real data distribution, the obtained hidden model is not really accurate. However, it is expected that by inverting the parameters of the global model, the hidden model can more or less capture some statistical characteristics of the global data distribution, and can thus provide direction-aware information towards I.I.D. cases. 
This hidden model can then fuzzily synthesize several mimic global I.I.D. data samples (sample features) with sampled softmax distribution of classes (i.e., decoded by hidden model), which will be further utilized to regularize the training process of each local model by a well-designed simple yet efficient bias-rectified objective. To simplify, we denote our model as \textit{Fed-FSNet}. 
Moreover, it should also be noted that since our proposed method is edge-agnostic and only synthesizes fuzzy mimic global I.I.D. data \textit{conditioned on only the global model}, we thus can still ensure the FL privacy \cite{bonawitz2017practical,so2021turbo}. 

In summary, the contributions of this paper are three-fold:
\begin{itemize}
\item We propose a novel FL training framework, dubbed \textit{Fed-FSNet}, involving a well-designed \textit{FSNet} to mitigate the Non-I.I.D. FL at-the-source. Our method can greatly narrow the gap between fully I.I.D. FL models.
\item The proposed \textit{FSNet} makes no inroads on the parameters/updates of local models nor analyzing individual local model outputs, thus can retain and ensure the privacy of the training process similar to conventional secure aggregation-based FL models.
\item Experimental results on several FL benchmarks verified the effectiveness of our model on the superior mitigation of Non-I.I.D FL.
\end{itemize}


\section{Related Work}
\label{Related_Work}

\subsection{Non-I.I.D. Data in FL}
Since different users usually have different habits and usage environments, the data samples generated on their devices are usually not independent and identically distributed (Non-I.I.D.) in real-world applications. 
Typically, during the FL training process, the Non-I.I.D. data means the differences between the data distributions $\mathcal{P}_{i}(x)$ and $\mathcal{P}_{j}(x)$ for any different devices $\mathcal{C}_{i}, \mathcal{C}_{j} \sim \mathcal{Q}$, where $\mathcal{Q}$ denotes the entire participating devices. Ideally, an I.I.D. (or at least quasi-I.I.D.) assumption of the training data is of crucial importance to ensure that the stochastic gradient is an unbiased estimate of the full gradients and thus makes the entire training convergence. In practice, it is however not realistic to assume that the local data on each device is always I.I.D. 
To mitigate the Non-I.I.D. issue in FL, \textit{MOCHA}~\cite{smith2017federated} proposed a multi-task learning framework to handle the statistical Non-I.I.D. challenge which is robust to practical systems issues. But this solution differs significantly from the previous work on FL. \textit{FedAvg}~\cite{mcmahan2017communication} has demonstrated that the iterative model averaging can work with certain non-I.I.D. data. \textit{FedAtt}~\cite{ji2019learning} proposed an attention mechanism in FL considering the contribution of each local model for aggregation and constructed a more generalized mobile keyboard suggestion model. More recently, \textit{AFL}~\cite{goetz2019active} proposed a simple and intuitive active device sampling scheme for FL training, with a probability conditioned on the current model and the data on the devices to offset the Non-I.I.D. Similarly, \textit{FAVOR}~\cite{wang2020optimizing} proposed a deep reinforcement learning-based experience-driven control framework that can intelligently sample devices to participate in each FL round to counterbalance the non-I.I.D. bias. Recently, knowledge distillation has been utilized to refine the server model aggregation using aggregated knowledge from heterogeneous users and obtained promising results, the representative methods include \textit{FedDF}~\cite{lin2020ensemble} and \textit{FedGEN}~\cite{zhu2021data}. 

\subsection{Inference Attacks as a Preliminary}
Recently, some generative models, e.g., auto-encoder (AE) \cite{vincent2010stacked,guo2019ee,chu2017stacked,guo2019adaptive} and generative adversarial network (GAN) \cite{goodfellow2014generative,isola2017image,karras2020analyzing}, and some optimization approaches are utilized for the inference attack on networks. 
With the ability to synthesize data samples of the target distribution \cite{vincent2010stacked,goodfellow2014generative,guo2019ams}, the generative models have proven to be efficient for inference attacks on ML models and thus can be used to recover privacy-sensitive data for certain purposes. 
For example, Hitaj~et~al.~\cite{hitaj2017deep} proposed an inference attack model against collaborative deep learning that allows the attacker to reconstruct sensitive information on the victim's device. \textit{mGAN-AI} \cite{wang2019beyond} incorporated a novel multi-task discrimination process on the reality, category, and identity of the target device for inferring user-level class representatives. 
Differently, \textit{DLG} \cite{zhu2019deep} applied an optimization method to match the gradients of the model on the malicious attacker and the gradients of the model trained on the target device to recover privacy-sensitive information. 

Motivated by inference attacks, we can also have the idea to make use of the related tools, especially the generative models to synthesize mimic I.I.D. data for participated devices, and mitigates the Non-I.I.D. FL at-the-source. Specifically, it is noticed that the above inference attack models usually require \textit{device-level information}, e.g., individual local model parameters/updates or outputs, to recover device-level data distribution, which however, naturally violates the FL privacy. Different from the above inference attack methods, we consider maintaining an edge-agnostic hidden model by training an encoder-decoder architecture conditioned on only the global model, to fuzzily synthesize mimic I.I.D. data in a global view. Our intuition is that although the synthesized mimic I.I.D. data is not fully accurate due to the absence of local model parameters/updates, it can still retain certain statistical characteristics of the overall global distribution that can provide direction-aware information towards I.I.D. cases. Such information is further utilized to regularize the FL training towards faster and better convergence.

\section{Methodology}
\label{Methodology}
We first give the problem definition of the FL training. Next, we introduce our proposed method and formulation in detail. More specifically, our motivation is that since the Non-I.I.D. brings about the biased estimation of each local model, we thus can make use of the mimic I.I.D. data (synthesized by the proposed \textit{FSNet}) to properly regularize the training process of each local model and offsets such bias at-the-source.

\subsection{Problem Definition}
\label{Problem_Definition}
We start by formalizing the FL training and then introduce our proposed method. Given $K$ edge devices (e.g., mobile phones) $\left \{ \mathcal{C}_{1}, \mathcal{C}_{2}, \cdots , \mathcal{C}_{K} \right \} \sim \mathcal{Q}$, where $\mathcal{Q}$ denotes the entire participated devices. Each $\mathcal{C}_{i}$ holds a private dataset belonging to $\left \{ \mathcal{D}_{1}, \mathcal{D}_{2}, \cdots , \mathcal{D}_{K} \right \}$, where the data distributions $\mathcal{P}_{i}(x)$ and $\mathcal{P}_{j}(x)$ for any different devices $\mathcal{C}_{i}, \mathcal{C}_{j} \sim \mathcal{Q}$ differ significantly. During the FL training, all participating devices are expected to collaboratively learn a shared global model. Without loss of generality, we consider the shared model to deal with a multi-class classification task, and thus each device trains a local model as:
\begin{equation}
\label{eq1}
\underset{\theta}{\min} \ \mathbb{E}_{\left (x, y\right )\in \mathcal{D}} \left [ -y \log \mathcal{P}_{\theta}\left (\hat{y} \mid x \right ) \right ],
\end{equation}
where $\mathcal{P}_{\theta}(\cdot)$ is the classifier with trainable parameter $\theta$, and $\hat{y}$ is the probability distribution (e.g., softmax outputs) that a data sample $x$ is predicated on all classes compared with the ground-truth $y$. 
Considering $K$ participated devices during the FL training, we can rewrite the objective (\ref{eq1}) as:
\begin{equation}
\label{eq2}
\underset{\theta_{1}, \theta_{2}, \cdots , \theta_{K}}{\min} \ \ \sum_{k=1}^{K} \mathbb{E}_{\left (x, y\right )\in \mathcal{D}_{k}} \left [ -y \log \mathcal{P}_{\theta_{k}}\left (\hat{y} \mid x \right ) \right ],
\end{equation}
where all $K$ devices are trained collaboratively for a shared global model, i.e., $\mathcal{P}_{\theta^{(g)}}(\cdot)$. 
Specifically, 
$\mathcal{P}_{\theta^{(g)}}(\cdot)$ is obtained by iteratively performing local update and model aggregation in each communication round as:
\begin{itemize}
\item \textit{Local Update:} Each participated device updates the local model on its private dataset using stochastic gradient descent (SGD) as $\theta_{k} \leftarrow  \theta_{k} - \eta \nabla \mathcal{P}_{\theta_{k}} (\mathcal{B}_{k})$, where $\mathcal{B}_k$ denotes the batch samples for one local iteration. 
\item \textit{Model Aggregation:} After the local update, the cloud server aggregates the updated local model parameters as $\theta^{(g)} = \sum_{k=1}^{K} \frac{N_{k}}{N}\theta_{k}$, where $N_{k}$ and $N$ denote the number of data samples in $\mathcal{D}_{k}$ and the total number of all datasets, respectively. The updated global model is then distributed to all participating devices and repeats the steps.
\end{itemize}

\begin{figure}[t]
  \centering
  \includegraphics[width=0.8\textwidth]{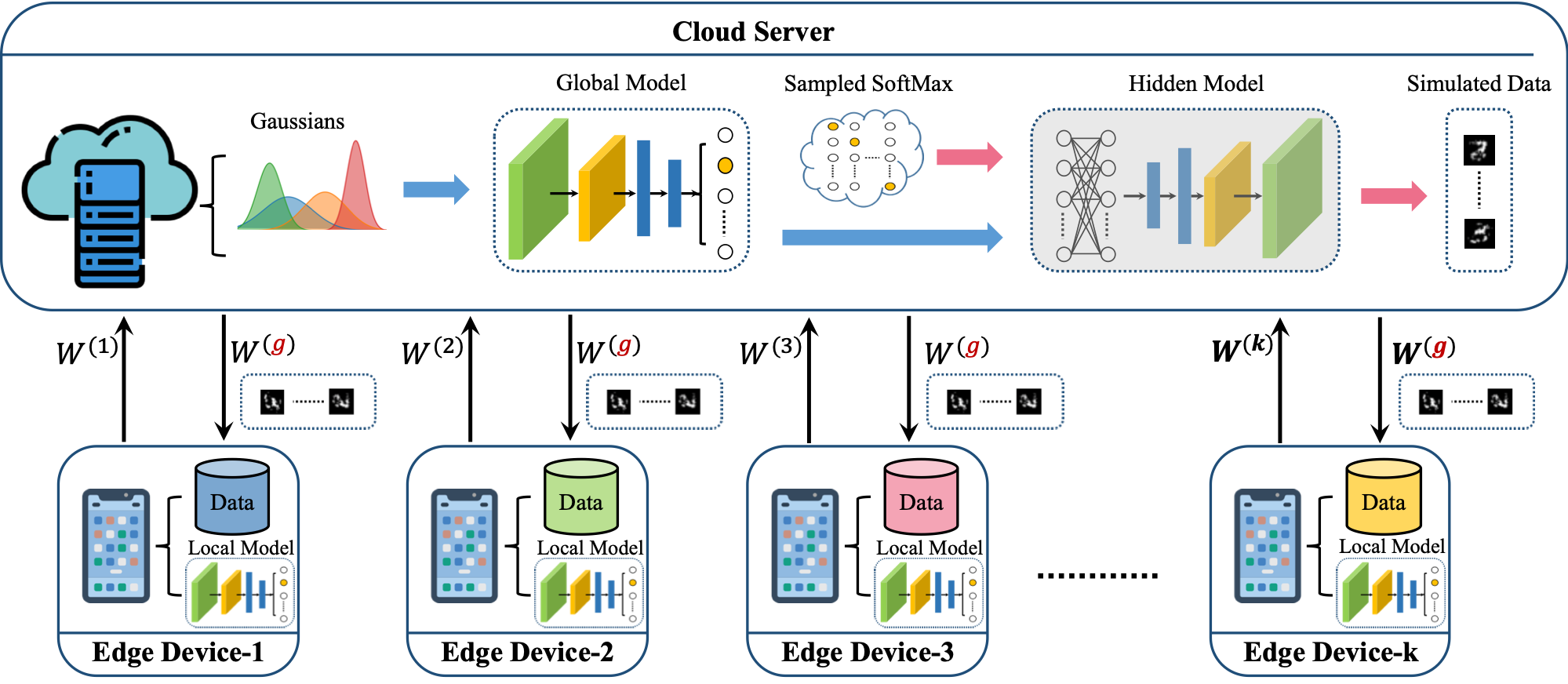}
  \caption{Illustration of the proposed FL training framework: an edge-agnostic hidden model is maintained in the cloud server to synthesize mimic I.I.D. data samples (sample features) conditioned on only the global model. These synthesized data will be offloaded along with the shared global model/parameters to each local device and further regularize the FL training.}
  \label{proposed}
\end{figure}

\subsection{Fuzzy Synthesizing Network (\textit{FSNet})}
\label{FSNet}
To synthesize mimic I.I.D. data, we introduce the fuzzy synthesizing network (\textit{FSNet}). As demonstrated in Figure~\ref{proposed}, our proposed FL training framework involving the \textit{FSNet} can be decomposed into three major components: 1) the FL base training pipeline which is similar to \textit{FedAvg}~\cite{mcmahan2017communication}; 2) the \textit{FSNet} that maintains an edge-agnostic hidden model in the cloud server conditioned on only the global model, which can be used to synthesize mimic I.I.D. data samples (sample features); and 3) the joint training scheme of the proposed \textit{Fed-FSNet} that offloads synthesized mimic I.I.D. samples (sample features) to each participated device and rectifies the bias caused by heterogeneous local data distribution, which can then mitigates the Non-I.I.D. FL at-the-source. 

\subsubsection{Training the \textit{FSNet}}
In our method, we use the encoder-decoder architecture to train the \textit{FSNet}, of which the shared global model acts as the encoder that takes data samples as inputs and maps them to the label space (i.e., normalized probability distribution such as softmax outputs), and a newly designed hidden model acts as the decoder to reconstruct the data samples. The training process can be described as:
\begin{equation}
\underset{\omega}{\min} \ \ \mathbb{E}_{{z} \in G_{s}} \left \| \mathcal{H}_{\omega}\left ( \mathcal{P}_{\theta^{(g)}}\left ( z \right ) \right ) - z \right \|_{2},	
\end{equation}
where $\mathcal{P}_{\theta^{(g)}}(\cdot)$ and $\mathcal{H}_{\omega}(\cdot)$ are the shared global model with fixed parameter $\theta^{(g)}$ and the hidden model with trainable parameter $\omega$, respectively. 
$\left \| \cdot \right \|_{2}$ is the $L_{2}$ distance that measures the difference between the input sample and the reconstructed counterpart. 
%
Specifically, $z$ denotes the samples sampled from a priori, e.g., Gaussian mixture with variational mixture coefficients ${\textstyle \sum_{k=1}^{K}\pi_{k}} \mathcal{N}(\mu_{k}, \Sigma_{k})$, where $\pi_{k}$ denotes the variational mixture coefficients. 

Notably, in our method, to better achieve the inversion of the shared global model, each local device is required to upload two simple scalars including its local mean $\mu_{k}$ and variance $\Sigma_{k}$ to form the Gaussian mixture in the cloud server, which will not violate the FL privacy.
Since there is no real data sample available in the cloud server for privacy preservation, we thus cannot have the observation on the real global distribution to train the hidden model $\mathcal{H}_{\omega}(\cdot)$ for an accurate reconstruction. 
However, we can expect that if the shared global model $\mathcal{P}_{\theta^{(g)}}(\cdot)$ is properly trained from all participated devices, it can certainly more or less retain some statistical characteristics of the global distribution in learned parameters $\theta^{(g)}$. 
Thus, in our method, when the fixed shared global model $\mathcal{P}_{\theta^{(g)}}(\cdot)$ and the hidden model $\mathcal{H}_{\omega}(\cdot)$ are trained in an \textit{end-to-end} manner with variational samplings from ${\textstyle \sum_{k=1}^{K}\pi_{k}} \mathcal{N}(\mu_{k}, \Sigma_{k})$, we can still expect $\mathcal{H}_{\omega}(\cdot)$ to be a less accurate inversion of the shared model $\mathcal{P}_{\theta^{(g)}}(\cdot)$ that can provide direction-aware information, e.g., gradients that towards the global one, to facilitate the FL convergence. 

Moreover, since the hidden model $\mathcal{H}_{\omega}(\cdot)$ is edge-agnostic and makes no inroads on the parameters/updates or outputs of each local model, it is thus can still ensure FL privacy. Hereby, we give a brief mathematical explanation of the privacy guarantee of our method. We define two scenarios $S_{1}, S_{2}$ as:
\begin{align}
&S_{1} := \sum_{k=1}^{N} \alpha_{k} \mathcal{P}_{\theta_{k}} \left ( \mathcal{D}_{k} \right ), \\
&S_{2} :=  \left \{\mathcal{P}_{\theta_{1}} \left ( \mathcal{D}_{1} \right ), \mathcal{P}_{\theta_{2}} \left ( \mathcal{D}_{2} \right ), \cdots , \mathcal{P}_{\theta_{K}} \left ( \mathcal{D}_{K} \right )  \right \},
\end{align}
where $S_{1}$ can be presented as the form of the shared edge-agnostic global model $\mathcal{P}_{\theta^{(g)}}(\cdot)$ and $S_{2}$ denotes each local model separately. We assume $\mathcal{P}_{\theta_{k}} \in \mathbb{R}^{d}$ and we can rewrite $S_{1}$ as the following matrix expression as:
\begin{align}
S_{1} := 
\mathbf{M} \begin{bmatrix}
\mathcal{P}_{\theta_{1}} \left ( \mathcal{D}_{1} \right )\\ 
\mathcal{P}_{\theta_{2}} \left ( \mathcal{D}_{2} \right )\\ 
\vdots \\ 
\mathcal{P}_{\theta_{N}} \left ( \mathcal{D}_{N} \right )
\end{bmatrix},
\end{align}
where $\mathbf{M}$ can be denoted as:
\begin{align}
\mathbf{M} = \left [ 
\begin{matrix}
\alpha_{1}  &  & \\
  & \ddots   & \\
  &  & \alpha_{1}
\end{matrix} 
\begin{matrix}
\alpha_{2}  &  & \\
  & \ddots   & \\
  &  & \alpha_{2}
\end{matrix}
\cdots 
\begin{matrix}
\alpha_{N}  &  & \\
  & \ddots   & \\
  &  & \alpha_{N}
\end{matrix} 
\right ].
\end{align}
Note:
\begin{equation}
\left\{\begin{matrix}
\begin{aligned}
& \mathrm{rank} \left ( \mathbf{M} \right ) = d \\ 
& \dim \left ( \ker \left ( \mathbf{M} \right ) \right ) = nd - d
\end{aligned}
\end{matrix}\right.	,
\end{equation}
for $n > 1$, $nd - d >  d$, it is obvious that $\mathbf{M} \vec{\mathcal{P}_{\theta}} = \sum_{k=1}^{N} \alpha_{k} \mathcal{P}_{\theta_{k}} \left (\mathcal{D}_{k} \right )$ has \textit{non-unique} solution which means that an accurate estimation of $S_{2}$ is impossible from the observation on only $S_{1}$. As a result, the proposed FL training framework can still preserve the privacy of each local device, which is similar to the secure aggregation-based FL models~\cite{bonawitz2017practical,so2021turbo}.

\subsubsection{Synthesizing Mimic I.I.D. Data}
Next, we can use the decoder part of the \textit{FSNet}, namely the hidden model $\mathcal{H}_{\omega}(\cdot)$, to synthesize mimic global I.I.D. data samples (sample features) and use them to regularize the training process of each local model. Specifically, we can know that the latent vector space of \textit{FSNet} is the normalized probability distribution such as the softmax outputs that represent the classes. Essentially, the softmax distribution ideally acts as both the output and input of the shared global model $\mathcal{P}_{\theta^{(g)}}(\cdot)$ and the hidden model $\mathcal{H}_{\omega}(\cdot)$, respectively, in an end-to-end manner. 
To synthesize mimic I.I.D. data samples (sample features), we can sample several representative softmax distributions from the latent vector space, and send them to the hidden model $\mathcal{H}_{\omega}(\cdot)$ to reconstruct them into the sample feature space. The synthesizing can be described as:
\begin{align}
{\mathcal{D}}' = \mathcal{H}_{\omega}\left ( \mathbf{S} = \begin{bmatrix}
C_{1} &\epsilon   &\cdots   &\epsilon \\ 
\epsilon &C_{2}  &\cdots   &\epsilon \\ 
\vdots  &\vdots   &\ddots   &\vdots  \\ 
\epsilon &\epsilon   &\cdots   &C_{n} 
\end{bmatrix} \right ),	
\end{align}
where ${\mathcal{D}}'$ is the synthesized dataset covering all global classes, and $\mathbf{S}$ is the sampled softmax distributions to be used as the input of the hidden model $\mathcal{H}_{\omega}(\cdot)$. More specifically, each column of $\mathbf{S}$ denotes a softmax output of a sample from target classes. For example, $\mathcal{H}_{\omega}(\left ( C_{1}, \varepsilon, \cdots, \varepsilon \right )^{\top})$ can condition the hidden model to synthesize a mimic data sample (sample features) for class ``$1$'' out of ``$n$'' classes, where $\sum \left ( C_{1}, \varepsilon, \cdots, \varepsilon \right )^{\top} = 1$ and $C_{1} \gg others$. It should be noted that since $\mathcal{H}_{\omega}(\cdot)$ is conditioned on $\mathcal{P}_{\theta^{(g)}}(\cdot)$ during training and conditioned on sampled $\mathbf{S}$ during synthesizing, the obtained mimic I.I.D. data samples (sample features) are usually not fully accurate. Nevertheless, this fuzzily synthesized dataset ${\mathcal{D}}'$ can still retain certain statistical characteristics of the overall global distribution that can provide direction-aware information, e.g., gradients that toward the global one, to facilitate the FL convergence. Such information can be used to regularize the training process and mitigates the Non-I.I.D. FL.

\subsection{Joint Training on \textit{Fed-FSNet}}
\label{Fed_FSNet}
With the mimic I.I.D. data samples ${x}' \in {\mathcal{D}}'$, we can then offload them (or extracted sample features) to each local device and train jointly with local data. Our intuition is that for each local dataset, i.e., $\left \{ \mathcal{D}_{1}, \mathcal{D}_{2}, \cdots , \mathcal{D}_{K} \right \}$, the local model is trained to classify its samples as accurate as possible and thus to retain and model the real local distribution. However, since the local data is Non-I.I.D. in each local device, the accurate modeling of each local dataset may eventually result in biased estimation of the global distribution and degrades the global model performance. To resolve the biased estimation, we can make use of the mimic global I.I.D. data, i.e., ${\mathcal{D}}'$, to force each local model to have the awareness of the global distribution as:
\begin{equation}
\label{eq10}
\begin{aligned}
\underset{\theta}{\min} \ \ 
&\underset{(i)}{\underbrace{\mathbb{E}_{\left (x, y\right )\in \mathcal{D}} \left [ -y \log \mathcal{P}_{\theta}\left (\hat{y} \mid x \right ) \right ]}} + \underset{(ii)}{\underbrace{\beta \mathbb{E}_{{x}' \in \mathcal{{D}'}} \left [ \mathcal{D_{KL}} \left ( \mathcal{U}\left ( \tilde{y} \right ) \parallel \mathcal{P}_{\theta}\left (\hat{y} \mid {x}' \right ) \right ) \right ]}},
\end{aligned}
\end{equation}
where term $(i)$ corresponds to the classification loss that mainly guides the model to accurately classify samples of each local device. Term $(ii)$ is the Non-I.I.D. regularizer that presents mimic global I.I.D. data, i.e., $\mathcal{{D}'}$, in each local model and forces them to shift the discrimination towards the global distribution. $\beta$ is a hyper-parameter that controls the balance of term $(ii)$.
More specifically, $\mathcal{D_{KL}}\left ( \cdot \right )$ calculates the Kullback-Leibler (KL) divergence, a.k.a. relative entropy, between the predicted probability distribution of the mimic global I.I.D. data $\mathcal{P}_{\theta}\left (\hat{y} \mid {x}' \right )$ and the uniform distribution $\mathcal{U}\left ( \tilde{y} \right )$. 
To minimize the KL divergence of term $(ii)$, we can force the classification results, i.e., probability distributions of all mimic global I.I.D. data samples belonging to $\mathcal{{D}'}$, to approximate the uniform distributions. Thus, the obtained model will result in the flat probability distributions of them which means that the local model has endowed with the awareness of the global distribution while maintaining the discrimination of real local distribution at the same time. We call Eq. (\ref{eq10}) the bias-rectified function for FL training, which resolves the biased estimation of local models and further makes the global aggregation more accurate in the cloud server. 
Similarly, considering $K$ participated devices during the FL training, we can rewrite it as:
\begin{equation}
\label{eq11}
\begin{aligned}
\underset{\theta_{1}, \theta_{2}, \cdots , \theta_{K}}{\min} \ \ 
&\mathbb{E}_{\left (x, y\right ) \in \mathcal{D}_{k}} \left [ -y \log \mathcal{P}_{\theta_{k}}\left (\hat{y} \mid x \right ) \right ] \\
&+ \beta \mathbb{E}_{{x}' \in \mathcal{{D}'}} \left [ \mathcal{D_{KL}} \left ( \mathcal{U}\left ( \tilde{y} \right ) \parallel \mathcal{P}_{\theta_{k}}\left (\hat{y} \mid {x}' \right ) \right ) \right ].
\end{aligned}
\end{equation}

The shared global model $\mathcal{P}_{\theta^{(g)}}(\cdot)$ is then obtained by iteratively performing local update and model aggregation in each communication round. 
To cooperatively optimize Eq. (\ref{eq11}) and obtain the shared global model, we construct an iterative update algorithm in Algorithm \ref{algorithm1}.
\begin{algorithm}[t]
Initialize shared global model $\mathcal{P}_{\theta^{(g)}}(\cdot)$ \;
Randomly select $m$ devices, offload the shared global model weight $\theta^{(g)}$ to each device\;
Local training on selected $m$ devices with Eq. (\ref{eq2}), obtain $\theta_{1}$, $\theta_{2}$, $\cdots$, $\theta_{m}$, and $\mu_{k}$ / $\Sigma_{k}$ pair \;
\Repeat
{\text{convergence}}
{

\For {$round = 1, 2, \cdots, T$}
{
Upload $\theta_{1}$, $\theta_{2}$, $\cdots$, $\theta_{m}$, and each $\mu_{k}$ / $\Sigma_{k}$ pair to the cloud server \;
Cloud server updates the shared global model weight as $\theta^{(g)} = \sum_{k=1}^{m} \frac{N_{k}}{N}\theta_{k}$ \;
Cloud server trains the hidden model $\mathcal{H}_{\omega}(\cdot)$ with Eq. (3) \;
Cloud server synthesizes the mimic I.I.D. data ${\mathcal{D}}'$ with Eq. (9) \;
Randomly select $m$ devices, offload the updated global model weight $\theta^{(g)}$ and mimic I.I.D. data ${\mathcal{D}}'$ (or extracted sample features) to each device \;
Local training on selected $m$ devices with Eq. (\ref{eq11}), obtain updated $\theta_{1}$, $\theta_{2}$, $\cdots$, $\theta_{m}$, and $\mu_{k}$ / $\Sigma_{k}$ pair.} 


}
\caption{Iterative Update}
\label{algorithm1}
\end{algorithm}

\section{Experiments}
\label{Experiments}

\subsection{Task and Dataset}
In our experiments, we implement the shared FL global model of the proposed \textit{Fed-FSNet} as the image classification task, and we evaluate the performance on three benchmark datasets adapted in the ``\textit{edge device FL training environment}'' including MNIST~\cite{lecun1998gradient}, Fashion-MNIST~\cite{xiao2017fashion}, and CIFAR-10~\cite{krizhevsky2009learning}.
For each evaluation, we set a total of 100 edge devices with one cloud server for the FL training. 
In our experiments, we consider an extreme Non-I.I.D. data scenario for participated devices. To prepare the Non-I.I.D. data for each dataset, we first split the train-set into 200 sorted partitions and then disorder them randomly. Afterward, each edge device is offloaded with two random partitions of the above train-set. For each dataset, its test-set is used for a global test after each communication round. 
We designed three simple model architectures that fit the on-device training Scenario for MNIST~\cite{lecun1998gradient}, CIFAR-10~\cite{krizhevsky2009learning}, and Fashion-MNIST~\cite{xiao2017fashion}, respectively.

\subsection{Competitor and Metric}

We compare the proposed \textit{Fed-FSNet} with representative FL models including \textit{FedAvg}~\cite{mcmahan2017communication}, \textit{AFL}~\cite{goetz2019active}, \textit{FedDF}~\cite{lin2020ensemble}, and \textit{FedGEN}~\cite{zhu2021data}. Among them, the \textit{FedAvg}~\cite{mcmahan2017communication} is the most standard FL training framework using averaging aggregation, which can be regarded as the baseline of FL. The other competitors adopt either data/devices sampling or improved aggregation strategies to mitigate the Non-I.I.D. issue.
The selection criteria for our competitors are: 1) recent work: all of these competitors are released in the most recent years; 2) non-I.I.D. oriented models: all of these competitors address the data non-I.I.D. issue in federated learning; and 3) competitiveness: they clearly represent the state-of-the-art. 
The prediction accuracy is selected as the evaluation metric. Specifically, during each communication round, we record two predicted results including the averaged accuracy across all participated devices, i.e., denoted as $Acc_{local}$, and the global accuracy, i.e., denoted as $Acc_{global}$. 
It is noticed that, in the experiments, the ``\textit{upper-bound}'' is set as the performance of \textit{FedAvg}~\cite{mcmahan2017communication} (most standard FL training framework using averaging aggregation) in a \textit{totally I.I.D. data setting} where each device shares full classes, and the ``\textit{lower-bound}'' is set as its performance in the generated extreme Non-I.I.D. data setting.

\subsection{Hyper-parameters}
In our experiments, we use stochastic gradient descent (SGD) to optimize the training. Each evaluation is trained with only 50 global epochs, and each local model is updated with 10 local epochs, respectively. The mini-batch size is set to 60 with a learning rate of 0.01. The hyper-parameter $\beta$ in Eqs. (10$\sim $11) is initially set to 1 and then decreased to one-tenth (1/10) in every 10 global rounds. During the training process, $\beta$ controls the balance of the Non-I.I.D. regularizer, i.e., term $(ii)$. The intuition is that at every beginning, each local device has very limited global information, thus a larger $\beta$ can force local models to shift their discrimination towards the global one. As the training goes on, $\beta$ trends to be decreased because more and more real global information is learned by the models.
As to the hyper-parameter $m$ in Algorithm \ref{algorithm1} that denotes the number of participated edge devices during the FL training, we set a fraction $\texttt{f}$=1.0 and $\texttt{f}$=0.1 denoting two scenarios where all 100 devices and only 10 randomly selected devices of each epoch are trained during the FL training.
Last, we force the sum of the variational mixture coefficients equals to 1 for the Gaussian mixture, i.e., ${\textstyle \sum_{i=1}^{K}} \pi_{k} = 1$.
%

\subsection{Evaluation}
\subsubsection{Prediction Results}
The comparison of three benchmark datasets is demonstrated in Table~\ref{prediction} and Table~\ref{results}.
Specifically, Table~\ref{prediction} compares the proposed \textit{Fed-FSNet} with the defined \textit{upper-bound} and \textit{lower-bound}, respectively. It can be observed from the results that our method surpasses the \textit{lower-bound} by a significant margin in the extreme Non-I.I.D. data setting. The global accuracy and averaged local accuracy of the proposed \textit{Fed-FSNet} achieve 88.2\% / 85.3\%, 61.7\% / 60.4\%, and 66.4\% / 64.1\% for MNIST \cite{lecun1998gradient}, CIFAR-10 \cite{krizhevsky2009learning}, and Fashion-MNIST \cite{xiao2017fashion} when $\texttt{f}$=1.0, respectively. 
Table~\ref{results} compares the proposed \textit{Fed-FSNet} with selected representative FL methods in fully Non-I.I.D. ($\texttt{f}$=0.1) setting. We can observe that our method constantly obtains the best performance for global accuracy and averaged local accuracy, respectively. 

In summary, the large margins with only 50 global epochs demonstrate the effectiveness of our method in two-fold: 1) our method can efficiently mitigate the Non-I.I.D. issue in FL training; and 2) our method is on-device friendly for the fewer training epochs and simple yet efficient architecture.

\begin{table*}[h]
\centering
\begin{threeparttable}
\caption{Prediction results compared with ``\textit{upper-bound}'' and ``\textit{lower-bound}'', respectively. The accuracy is reported in the averaged accuracy. $Acc_{global}$ (\%) denotes the averaged accuracy across all participated devices, and $Acc_{local}$ (\%) denotes the global accuracy. The \textit{upper-bound} is set as the performance of \textit{FedAvg} in a totally I.I.D. data setting, and the \textit{lower-bound} is set as the performance of \textit{FedAvg} in the generated extreme Non-I.I.D. data setting.}
\label{prediction}
\setlength{\tabcolsep}{0.57mm}{    
\begin{tabular}{lcccccccc}  
\toprule  
\multirow{2}{*}{Method}
&\multicolumn{2}{c}{MNIST}
&\multicolumn{2}{c}{CIFAR-10}
&\multicolumn{2}{c}{Fashion-MNIST}\cr   
\cmidrule(lr){2-3} 
\cmidrule(lr){4-5} 
\cmidrule(lr){6-7}
   &$Acc_{global}$  &$Acc_{local}$              &$Acc_{global}$   &$Acc_{local}$       &$Acc_{global}$   &$Acc_{local}$    \cr  
\midrule
\textit{Upper-bound} / I.I.D. ($\texttt{f}$=0.1)       &97.6        &97.3       &72.6     &71.9       &76.4    &74.2      \cr
\textit{Upper-bound} / I.I.D. ($\texttt{f}$=1.0)       &97.9         &97.4       &75.6     &74.3       &79.3     &77.1     \cr
\hline
\textit{Lower-bound} / Non-I.I.D. ($\texttt{f}$=0.1)       &37.1         &49.8       &21.1     &27.6   &25.4     &32.7             \cr
\textit{Lower-bound} / Non-I.I.D. ($\texttt{f}$=1.0)       &42.1         &35.2       &31.8     &20.0       &38.6     &23.3     \cr
\hline
\textbf{\textit{Fed-FSNet} (ours)} / Non-I.I.D. ($\texttt{f}$=0.1)       &\textbf{81.5}         &\textbf{85.1}        &\textbf{56.3}      &\textbf{58.9}        &\textbf{59.7}      &\textbf{61.4}      \cr
\textbf{\textit{Fed-FSNet} (ours)} / Non-I.I.D. ($\texttt{f}$=1.0)       &\textbf{88.2}          &\textbf{85.3}        &\textbf{61.7}      &\textbf{60.4}        &\textbf{66.4}      &\textbf{64.1}     \cr
\bottomrule  
\end{tabular}
} 
\end{threeparttable}
\end{table*}

\begin{table*}[h]
\centering
\begin{threeparttable}  
\caption{Prediction Results compared with representative models in Non-I.I.D. ($\texttt{f}$=0.1) setting. The accuracy is reported in the averaged accuracy. $Acc_{global}$ (\%) denotes the averaged accuracy across all participated devices, and $Acc_{local}$ (\%) denotes the global accuracy.}  
\label{results}
\setlength{\tabcolsep}{2.7mm}{    
\begin{tabular}{lcccccccc}  
\toprule  
\multirow{2}{*}{Method}
&\multicolumn{2}{c}{MNIST}
&\multicolumn{2}{c}{CIFAR-10}
&\multicolumn{2}{c}{Fashion-MNIST}\cr  
\cmidrule(lr){2-3} 
\cmidrule(lr){4-5} 
\cmidrule(lr){6-7}
   &$Acc_{global}$  &$Acc_{local}$              &$Acc_{global}$   &$Acc_{local}$       &$Acc_{global}$   &$Acc_{local}$    \cr  
\midrule
\textit{FedAvg} \cite{mcmahan2017communication}       &37.1         &49.8       &21.1     &27.6 &25.4     &32.7            \cr
\textit{AFL} \cite{goetz2019active}       &54.2         &59.7       &33.8     &42.2   &37.9     &49.8            \cr
\textit{FedDF} \cite{lin2020ensemble}       &74.5         &83.3       &49.7     &52.5   &51.7     &57.3            \cr
\textit{FedGEN}~\cite{zhu2021data}       &75.1         &83.9       &53.1     &54.0   &52.8     &57.6            \cr
\textbf{\textit{Fed-FSNet} (ours)}       &\textbf{81.5}         &\textbf{85.1}        &\textbf{56.3}      &\textbf{58.9}        &\textbf{59.7}      &\textbf{61.4}     \cr
\bottomrule  
\end{tabular}
} 
\end{threeparttable}
\end{table*}

\subsubsection{Training Analysis}
\begin{figure*}[h]
  \centering
  \subfigure[Global Accuracy (\texttt{f}=1.0): $Acc_{global}$]{
    \label{fig:subfig:a} 
    \includegraphics[width=1.77in]{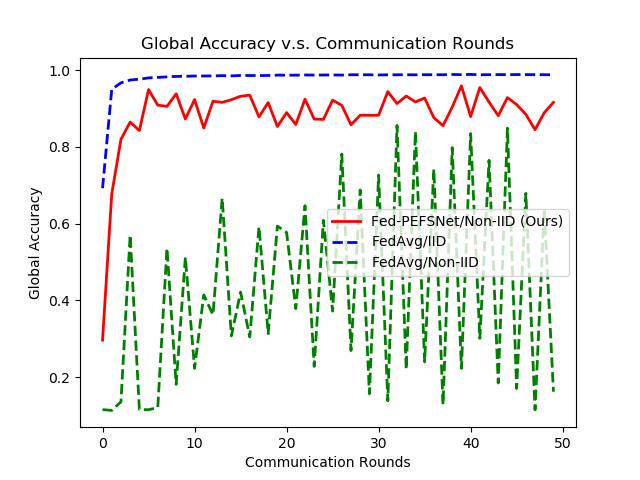}}
  \subfigure[Local Accuracy (\texttt{f}=1.0): $Acc_{local}$]{
    \label{fig:subfig:b} 
    \includegraphics[width=1.77in]{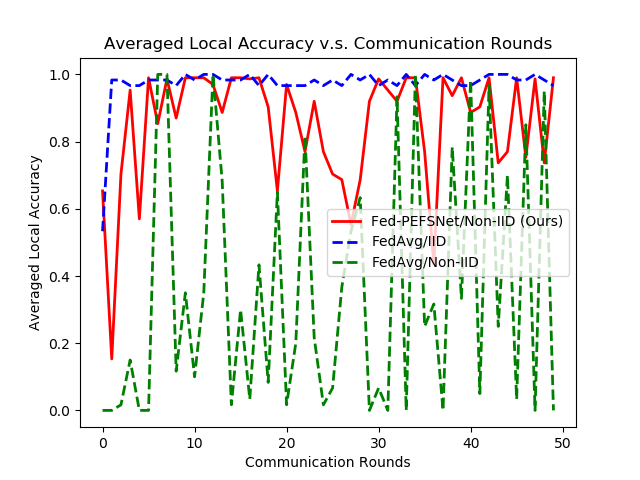}}
\subfigure[Training Loss (\texttt{f}=1.0)]{
    \label{fig:subfig:c} 
    \includegraphics[width=1.77in]{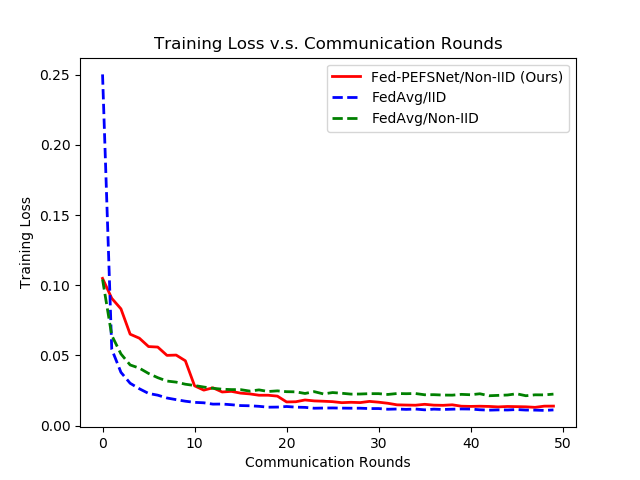}}

  \subfigure[Global Accuracy (\texttt{f}=0.1): $Acc_{global}$]{
    \label{fig:subfig:d} 
    \includegraphics[width=1.77in]{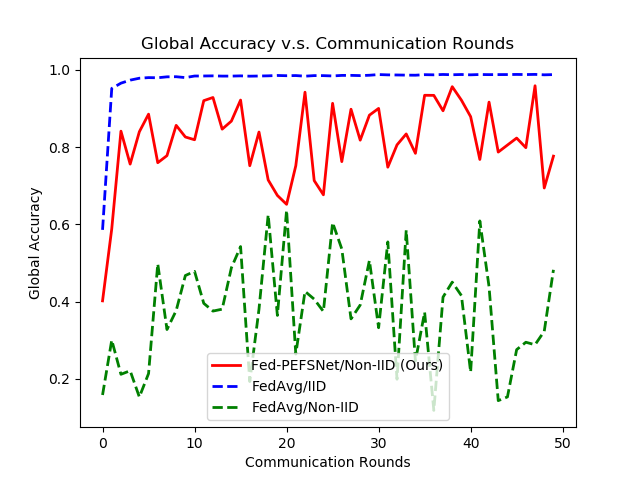}}
  \subfigure[Local Accuracy (\texttt{f}=0.1): $Acc_{local}$]{
    \label{fig:subfig:e} 
    \includegraphics[width=1.77in]{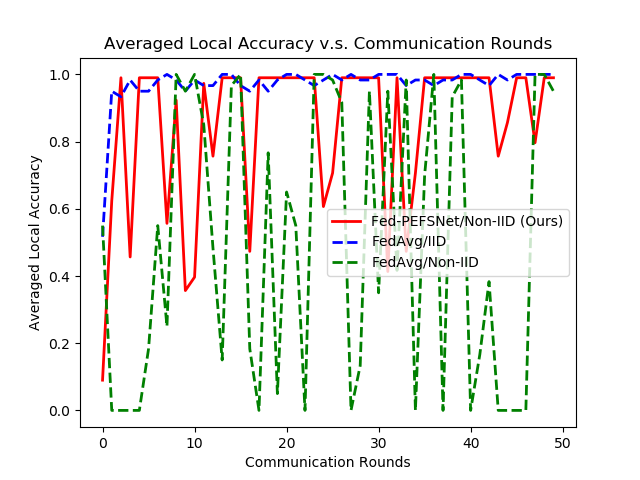}}
\subfigure[Training Loss (\texttt{f}=0.1)]{
    \label{fig:subfig:f} 
    \includegraphics[width=1.77in]{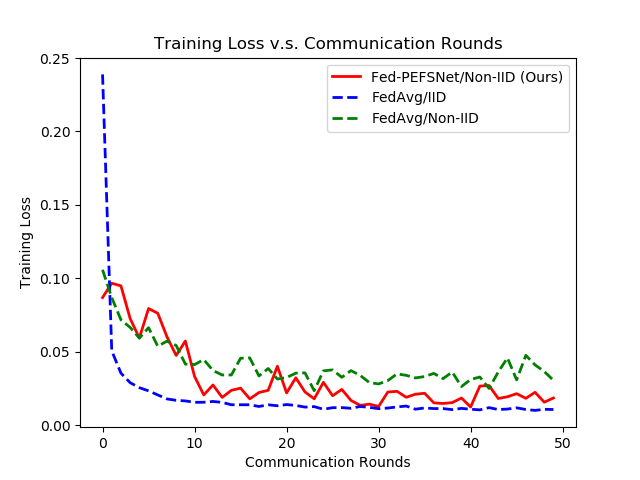}}
\caption{The training analysis of our proposed method. 
The red curve denotes our method with Non-I.I.D. settings, and the blue and green curves denote the upper bound (totally I.I.D.) and lower bound (Non-I.I.D.) of the baseline, respectively. 
}
  \label{visualization}
\end{figure*}

We visualize the training process of the proposed \textit{Fed-FSNet} with Non-I.I.D. settings on MNIST~\cite{lecun1998gradient} in Figure~\ref{visualization}. It can be observed from the results that the training process of our method (denoted as the red curve) is more stable and convergences much faster and better than the conventional Non-I.I.D. FL (denoted as green curve) by fully making use of the synthesized mimic I.I.D. data. 
It should be noted that, although our method still has a small margin from the totally I.I.D. FL (denoted as the blue curve), it greatly narrows the gap between non-I.I.D. and I.I.D., making it applicable for us to mitigate the Non-I.I.D. problem in FL and facilitates the implementation of practical applications.

\subsection{Discussion on Communication Overhead}
In our experiments, since the mimic dataset ${\mathcal{D}}'$ contains very few, i.e., only 60 synthesized samples (sample features) set in our experiment, it can result in a very limited impact on the communication overhead. 
Alternatively, if larger data, e.g., a CT volume, is utilized in real-world applications, the synthesized samples may increase exponentially. To solve this limitation, our solution is to deploy a pre-trained feature extractor in the cloud server and only transmit the extracted feature vector (far smaller) to each participating device, which can significantly reduce the communication overhead. 
Besides, since the proposed method requires fewer training epochs and simpler model architecture, thus it will be more suitable for on-device learning where the edge devices are usually equipped with less computing capabilities and fewer memories.

\section{Conclusion}
\label{Conclusion}
In this paper, we proposed a novel FL training framework involving a privacy ensured fuzzy synthesizing network (\textit{Fed-FSNet}) to mitigate the Non-I.I.D. FL at-the-source. Our method narrows the gap between the non-I.I.D. and I.I.D. FLs with a large margin. 
Experimental results on FL benchmarks verified the effectiveness of our model. 
In the future, we have two research directions to further improve the Non-I.I.D. FL. The first investigates more efficient and generalized methods to synthesize Non-I.I.D. data, and the second will focus on more privacy-preserved methods.

\bibliographystyle{unsrt}
\bibliography{my}

\end{document}